\documentclass[prd,superscriptaddress,amsfonts,amssymb,amsmath,showpacs]{revtex4-2}
\usepackage{bm}
\usepackage{amsfonts}
\usepackage{latexsym}
\usepackage{graphicx}
\usepackage{amsmath}
\usepackage{palatino}
\usepackage{mathpazo}
\usepackage{natbib}
\usepackage{textcomp}
\usepackage{booktabs}
\usepackage{dcolumn}
\usepackage{booktabs}
\usepackage{multirow}
\usepackage{hyperref}
\hypersetup{colorlinks,citecolor=red}
\usepackage{amsmath}
\usepackage{xcolor}
\usepackage{orcidlink}
\usepackage[caption=false]{subfig}
\usepackage{commath}
\usepackage{lmodern}

\usepackage{anyfontsize}
\usepackage{amssymb}
\usepackage{amsthm}
\usepackage{amsmath}
\usepackage{lipsum}
\usepackage{float}
\usepackage{orcidlink}

\def\jnl@style{\it}
\def\aaref@jnl#1{{\jnl@style#1}}

\def\aaref@jnl#1{{\jnl@style#1}}

\def\aj{\aaref@jnl{AJ}}                   
\def\apj{\aaref@jnl{ApJ}}                 
\def\apjl{\aaref@jnl{ApJ}}                
\def\apjs{\aaref@jnl{ApJS}}               
\def\apss{\aaref@jnl{Ap\&SS}}             
\def\aap{\aaref@jnl{A\&A}}                
\def\aapr{\aaref@jnl{A\&A~Rev.}}          
\def\aaps{\aaref@jnl{A\&AS}}              
\def\mnras{\aaref@jnl{Mon.~Not.~Roy.~Astron.~Soc.}}             
\def\prd{\aaref@jnl{Phys.~Rev.~D}}        
\def\prc{\aaref@jnl{Phys.~Rev.~C}}  
\def\prl{\aaref@jnl{Phys.~Rev.~Lett.}}    
\def\qjras{\aaref@jnl{QJRAS}}             
\def\skytel{\aaref@jnl{S\&T}}             
\def\ssr{\aaref@jnl{Space~Sci.~Rev.}}     
\def\zap{\aaref@jnl{ZAp}}                 
\def\nat{\aaref@jnl{Nature}}              
\def\aplett{\aaref@jnl{Astrophys.~Lett.}} 
\def\apspr{\aaref@jnl{Astrophys.~Space~Phys.~Res.}} 
\def\physrep{\aaref@jnl{Phys.~Rep.}}      
\def\physscr{\aaref@jnl{Phys.~Scr}}       
\def\commat{\aaref@jnl{Comm.~Math.~Phys.}}              
\def\science{\aaref@jnl{Science}}               
\def\cqg{\aaref@jnl{Classical Quant.~Grav.}}            
\def\jpcs{\aaref@jnl{JPCS}}                                     
\def\ijmpd{\aaref@jnl{Int.~J.~Mod.~Phys.~D}}                    
\def\grg{\aaref@jnl{Gen.~Relat.~Gravit.}}               
\def\rpp{\aaref@jnl{Rep.~Prog.~Phys.}}          
\def\npa{\aaref@jnl{Nucl.~Phys.~A}}        
\def\lrr{\aaref@jnl{Living Rev.~Rel.}}                   
\def\jcap{\aaref@jnl{J.~Cosmology Astropart.~Phys.}}    
\def\rmp{\aaref@jnl{Rev.~Mod.~Phys.}}   
\def\epjc{\aaref@jnl{Eur.~Phys.~J.~C}}

\allowdisplaybreaks[1]

\begin{document}

\color{black}       

\title{Cosmological Dynamics in $f(R,L_m,T)$ Modified Gravity}

\author{V. A. Kshirsagar}
\email{kvitthal99@gmail.com}
\affiliation{Shri Jagdishprasad Jhabarmal Tibrewala University, Vidhya Nagari, Churu-Jhunjhunu Road, Chudela, Jhunjhunu, Rajasthan 333010}
\author{A. S. Agrawal\orcidlink{0000-0003-4976-8769}}
\email{asagrawal.sbas@jspmuni.ac.in}
\affiliation{Department of Mathematics, School of Computational Sciences, JSPM University Pune-412207, India}
\author{S. A. Kadam\orcidlink{0000-0002-2799-7870}}
\email{sak.scos@jspmuni.ac.in;
\\k.siddheshwar47@gmail.com}
\affiliation{Department of Mathematics, School of Computational Sciences, JSPM University Pune-412207, India}
\author{Vishwajeet S. Goswami}
\email{vishwajeetgoswami.math@gmail.com}
\affiliation{Shri Jagdishprasad Jhabarmal Tibrewala University, Vidhya Nagari, Churu-Jhunjhunu Road, Chudela, Jhunjhunu, Rajasthan 333010}

\date{\today}

\begin{abstract}
\textbf{Abstract}: In this paper, we investigate the accelerating phase of the Universe within the context of $f(R,L_m,T)$ gravity theory, where $R$, $L_m$, and $T$ represent the Ricci scalar, matter Lagrangian, and the trace of the energy-momentum tensor, respectively. We focus on a particular form of modified gravity defined by $f(R,L_m,T) = R - \mu L_m T - \gamma$, with $\mu$ and $\gamma$ being positive constants. The matter sector is characterized by the Lagrangian density $L_m = -\rho$, where $\rho$ denotes the energy density of the cosmological fluid. We conduct an in-depth examination of the model using phase space analysis, thoroughly evaluating the evolution of cosmological solutions with dynamical system techniques.  The results is illustrated through graphs in the phase space, the characteristics of critical points and the stable attractors within the proposed modified gravity $f(R,L_m,T)$ cosmological framework. We investigate the transition from the initial decelerating phase of the universe to its current accelerating phase. The behaviour of the EoS, deceleration parameter with the appropriate initial conditions have been investigated.
\end{abstract}

\maketitle
\section{Introduction}\label{Introduction}
The investigation of alternative gravitational theories, especially those that incorporate the coupling between matter and geometry, is motivated by the limitations of the $\Lambda$CDM model. Although $\Lambda$CDM has shown a robust alignment with with observational data including findings from Type Ia supernovae \cite{Riess_1998, Perlmutter_1999}, the Planck satellite \cite{Planck:2018vyg}, and the Sloan Digital Sky Survey (SDSS) \cite{Tegmark_2004}. These data studies concludes the current Universe is experiencing an accelerating expansion of the Universe. The General theory of Relativity (GR) does not explains this accelerating phenomena with negative pressure \cite{Carroll:1991mt}. As a result, researchers in the field have explored two main approaches, either developing alternative theories to GR or formulating modifications to the GR. A key development in this field was Einstein's addition of the cosmological constant $\Lambda$ to his field equations \cite{Einstein:1917ce}. This idea subsequently evolved into the $\Lambda$CDM model, which effectively describes cosmic acceleration by incorporating dark matter and dark energy \cite{Peebles:2002gy, Planck:2018vyg}. Nonetheless, despite its empirical successes, $\Lambda$CDM faces theoretical issues that remain unresolved mainly regarding the discrepancy in observational and current value of cosmological constant $\Lambda$ \cite{RevModPhys_61_1, Bull_2016}.

To explore the phenomenon of cosmic acceleration, various alternative models have been proposed, including numerous dark energy theories and modifications to GR \cite{COPELAND_2006, Clifton_2012}. Some studies utilize thermodynamic methods to analyze the progression of the deceleration parameter \cite{Pavon:2012qn}, while others examine the effects of additional barotropic fluids \cite{Visser:2003vq}. Furthermore, cosmographic methods \cite{Visser:2003vq}, holographic models of dark energy \cite{Zhang:2005yz}, and analyses of curvature eigenvalues \cite{Nojiri:2006ri} have also been explored. Intermediate redshift calibration techniques have offered tools for evaluating cosmic correlations \cite{Lusso:2016vmu}. One of the initial modifications to GR is $f(R)$ gravity theory which redefines the Einstein-Hilbert action by replacing the Ricci scalar $R$ with a general function $f(R)$. This approach provides a purely geometric foundation for late-time cosmic acceleration without requiring dark energy assumptions. Although the theory coincides with GR in the case of $f(R) = R$, the addition of higher-order terms allows to study more complex cosmological behavior \cite{Sotiriou_2010, De_Felice_2010}. Recently, theories involving matter-geometry interactions have gained interest, where gravitational modifications arise from the interaction between curvature and matter. In particular, the $f(R,T)$ \cite{Harko_2011} and $f(R,L_m)$ \cite{Harko_2010} . The $f(R,T)$ framework incorporates both the Ricci scalar $R$ and the trace of the energy-momentum tensor $T$, introducing a correlation that enables non-conserved energy-momentum. This has substantial implications for cosmological dynamics, especially when considering imperfect fluids \cite{Harko_2011}. The $f(R,L_m)$ theory \cite{Harko_2010} expands the gravitational Lagrangian to include dependence on the matter Lagrangian $L_m$ in addition to $R$, allowing for direct interactions between matter and geometry. This model can produce extra forces acting on massive particles and is particularly beneficial for simulating interactions within the dark sector. The review outlines the developmental path of the hybrid metric-Palatini gravity framework, highlighting its consistency with local gravitational tests and its capability to tackle both cosmological and astrophysical issues. Importantly, this framework offers a unified theoretical structure that can clarify the mysteries surrounding dark energy and dark matter \cite{Harko_2012, Capozziello_2015, Carloni_2015}. These theoretical frameworks have been extensively investigated within cosmological and astrophysical environments.

Haghani and Harko \cite{Haghani_2021} have recently introduced an extended class of gravity theories characterized by a non-minimal interaction between matter and geometry, referred to as $f(R,L_m,T)$ gravity. Further examinations of hydrostatic equilibrium in compact stellar bodies within this framework have shown that the coupling between matter and geometry significantly influences the distributions of pressure and density \cite{Fortunato:2024ulg, Mota_2024}. Additionally, solutions related to wormholes were proposed in \cite{Moraes_2024, Loewer_2024}. The primary objective of this framework is to unify various modified gravity theories under Riemannian geometry \cite{Harko_2014, Capozziello_2011, Nojiri_2011}. Although alternative gravity models such as $f(R)$, $f(T)$, and $f(Q)$ have demonstrated potential in addressing cosmic acceleration phenomena \cite{Capozziello_2011, Nojiri_2011, Bamba_2012}, this study focuses on the investigating  less explored modified gravity $f(R,L_m,T)$ formalism to investigate cosmological dynamics of the Universe.

In this paper, we intend to examine the dynamics of a flat Friedmann-Lemaître-Robertson-Walker (FLRW) universe within the framework of $f(R, L_m, T)$ gravity. The choice of a flat FLRW geometry is supported by observational evidence from the Planck mission, which suggests that the universe’s spatial curvature is nearly zero \cite{Planck:2018vyg}. By concentrating on flat spacetime, we are able to distinguish the effects of the modified gravity framework from complications arising from curvature. Our approach combines theoretical modeling compatible to observational constraints, creating a robust foundation for uncovering the cosmological implications of the matter-geometry coupling. Specifically, we investigate whether the chosen formulation of $f(R, L_m, T)$ gravity can adequately explain the observed cosmic acceleration without requiring a cosmological constant or the introduction of exotic matter components. Additionally, we analyze the behavior of the effective equation of state and the deceleration parameter to assess the model’s alignment with existing observational data studies.

This paper is organized into five sections, Section \ref{Introduction} provides an introduction and a general overview of the research problem, Section \ref{sec:cwfrl} presents the theoretical formulation of $f(R,L_m,T)$ gravity, outlining its fundamental framework and principles. In Section \ref{sec:cwfrl3}, we employ the perfect fluid stress–energy tensor to derive the corresponding field equations for a spatially flat FLRW universe and subsequently obtain cosmological solutions for the specific model $f(R,L_m,T) = R - \mu L_m T - \gamma$ \cite{Haghani_2021} Section \ref{sec:cwfrl2} is devoted to the dynamical systems analysis, where we reformulate the field equations into autonomous equations, discuss the associated constraints, and analyze the critical points and phase–plane behavior of the system. Finally, Section \ref{conclusion} summarizes the main findings, highlights the physical implications, and outlines possible directions for future research.

\section{Basic formalism of $f(R,L_m,T)$ gravity}
\label{sec:cwfrl}

The action equation of $f(R,L_m,T)$ gravity can be presented as \cite{Haghani_2021}, 
\begin{equation}\label{frl1}
    S=\int\left[ L_m  +  f(R,L_m, T) \right]\sqrt{-g} d^4x,
\end{equation}
The $g$ determinant of the metric tensor is denoted. The term $f(R,L_m,T)$ signifies a general function that is a function of the Ricci scalar $R$, the trace of the energy-momentum tensor $T_{ij}$ represented as $T$, and the matter Lagrangian density $L_m$. This function $f(R,L_m,T)$ embodies a comprehensive framework that expands upon both $f(R,L_m)$ and $f(R,T)$ gravity theories, incorporating a non-minimal interaction between matter and geometry. This formulation accommodates a wide variety of matter fields and leads to non-trivial modifications of general relativity. Such theoretical developments provide strong motivation for exploring cosmological models in this extended gravitational context. We work in natural units where $c=1$ and the gravitational constant is rescaled such that $8\pi G_N=1$, which will be assumed \cite{Moraes_2024}. The energy-momentum tensor for normal (baryonic) matter is conventionally defined by the relation \cite{LandauLifshitz1980},
\[
 -\frac{2}{\sqrt{-g}} \frac{\delta (\sqrt{-g} \, L_m)}{\delta g^{ij}}=T_{ij},
\] and its trace is given by \( T = g^{ij} T_{ij} \). We consider the \( \mathcal{L}_m \) matter Lagrangian density depends only on the \( g_{ij} \) component of metric density, and not depend on its derivatives, then the stress-energy tensor simplifies to,
\begin{equation}
 g_{ij} L_m - 2 \frac{\partial L_m}{\partial g^{ij}}= T_{ij} 
 \end{equation} 

The variation of eq. (\ref{frl1}) with respect to the  \( g^{ij} \) metric tensor can be obtain,
\begin{equation}
\delta S=\int \left[ f_T \frac{\delta T}{\delta g^{ij}} \delta g^{ij} +f_R\delta R+ f_{L_m} \frac{\delta {L_m}} {\delta g^{ij}} \delta g^{ij} - 8 \pi T_{ij} \delta g^{ij}-\frac{1}{2} g_{ij} f \delta g^{ij}\right] \sqrt{-g} d^4x,
\label{eq:3}\end{equation}

Where we apply the notation respectively $f_R \equiv \frac{\partial f}{\partial R}, \quad
f_T \equiv \frac{\partial f}{\partial T}, \quad
f_{L_m} \equiv \frac{\partial f}{\partial L_m}$, 

The Ricci scalar $R$ is defined as the reduction of the Ricci tensor $R_{ij}$ with the metric tensor $g^{ij}$ is given by \begin{equation}\label{frl2}
R=g^{ij}R_{ij},
\end{equation}

The variation of (\ref{frl2}) can be obtained using the Palatini identity 
\begin{equation}
R_{ij} \, \delta g^{ij} + g^{ij} \delta R_{ij} =\delta (g^{ij} R_{ij}) = \delta R 
\end{equation}
\begin{equation}
\delta (g^{ij} R_{ij}) = R_{ij} \, \delta g^{ij} + g^{ij} \left( \nabla_k \delta \Gamma^k_{ij} - \nabla_j \delta \Gamma^k_{ik} \right) = \delta R 
\end{equation}

where \( \nabla_k \) means covariant derivative with respect to Levi-Civita connection related with the metric \( g \), while the $\Gamma^k_{ij}$ christoffel symbols are associated with metric. The variation of the Christoffel symbols with respect to the metric tensor is given by \cite{Harko_2011},
\begin{equation}
\frac{1}{2} g^{k\alpha} \left( \nabla_{i} \delta g_{j\alpha} + \nabla_{j} \delta g_{\alpha i} - \nabla_{\alpha} \delta g_{ij} \right)=\delta \Gamma^{k}_{ij}
\end{equation}

The expression resulting from the variation of the Ricci scalar is given by,
\begin{equation}
\left[R_{ij} + g_{ij}  \Box - \nabla_i \nabla_j \right]\delta g^{ij}=\delta R
\end{equation}
The variation of trace $T = T^{i}_{i}$ of matter energy--momentum tensor $T_{ij}$ with respect to the metric tensor $g_{ij}$ is expressed as
\begin{equation}
T_{ij} + \Theta_{ij} = \frac{\delta (g^{\alpha\beta} T_{\alpha\beta})}{\delta g^{ij}} 
\end{equation}
where the additional tensor \( \Theta_{ij} \) is defined by,
\begin{equation}
\Theta_{ij} \equiv g^{\alpha\beta} \frac{\delta T_{\alpha\beta}}{\delta g^{ij}} = L_m g_{ij} - 2 T_{ij} - \tau_{ij}    
\end{equation}

with the term \( \tau_{ij} \) representing the second variation of the matter Lagrangian density, given by,
\begin{equation}
\tau_{ij} = 2 g^{\alpha\beta} \frac{\partial^2 L_m}{\partial g^{ij} \, \partial g^{\alpha\beta}} \end{equation}

Using these relations in the variation of the action (as in Eq. \eqref{eq:3}), and applying integration by parts, the field equations corresponding to the $f(R,L_m,T)$ theory of modified gravity is expressed as,
\begin{equation}
\left(R_{ij} + g_{ij} \Box - \nabla_i \nabla_j \right) f_R - \frac{1}{2} f g_{ij}
= 8\pi T_{ij} + \frac{1}{2} \left(f_{L_m} + 2f_T \right)\left( T_{ij} - L_m g_{ij} \right) + f_T \, \tau_{ij}
\end{equation}
The covariant derivative is $\nabla_i$ and the D’Alembert operator is $\square=g^{ij} \nabla_i \nabla_j$. The derivatives of the function \( f \) are defined as,
\[
f_R \equiv \frac{\partial f}{\partial R}, \quad
f_{L_m} \equiv \frac{\partial f}{\partial L_m}, \quad
f_T \equiv \frac{\partial f}{\partial T}, \quad
\tau_{ij} \equiv 2 g^{\alpha\beta} \frac{\partial^2 L_m}{\partial g^{ij} \partial g^{\alpha\beta}}
\]

For a perfect fluid described by the matter Lagrangian $L_m = -\rho$, as well as for a scalar field with Lagrangian $L_m = -\frac{1}{2} \partial_i \phi \, \partial^i \phi + V(\phi)$, the additional tensor $\tau_{ij}$  turns out to be zero \cite{Haghani_2021}. Therefore we get the field equations of $f(R,L_m,T)$ gravity model which is,
\begin{equation}
\left(R_{ij} + g_{ij} \Box - \nabla_i \nabla_j \right) f_R - \frac{1}{2} f g_{ij}
= 8\pi T_{ij} + \frac{1}{2} \left(f_{L_m} + 2f_T \right)\left( T_{ij} - L_m g_{ij} \right)
\label{eq:13}\end{equation}
\section{Model and Basic Equations}\label{sec:cwfrl3}

We utilize the flat FLRW metric to model a spatially homogeneous and isotropic universe. The line element for this space--time is expressed as \cite{Ryden:1970vsj},
\begin{equation}
ds^2 = -dt^2 + a^2(t) \left( dx^2 + dy^2 + dz^2 \right)
\label{eq:14}\end{equation} 
where $a(t)$ represents the scale factor that procedures how the size of the universe changes with cosmic time $t$. In the context of metric defined in Eq. \eqref{eq:14}, the nonzero Christoffel symbols are given by,
\[\Gamma^0_{ij} = -\frac{1}{2} g^{00} \frac{\partial g_{ij}}{\partial x^0},\]
\[\Gamma^k_{0j} = \Gamma^k_{j0} = \frac{1}{2} g^{k\lambda} \frac{\partial g_{j\lambda}}{\partial x^0}\]
where $i,j,k=0,1,2,3$
The non-vanishing modules of the Ricci curvature tensor can be computed as,
\[R^1_1 = R^2_2 = R^3_3 = \frac{\ddot{a}}{a} + 2\left( \frac{\dot{a}}{a} \right)^2,\]
\[R^4_4 = 3\frac{\ddot{a}}{a}\]

As a result, the Ricci scalar $R$, corresponding to the line element in Eq. \eqref{eq:14}, takes the form,
\begin{equation}
R = 6\frac{\ddot{a}}{a} + 6\left( \frac{\dot{a}}{a} \right)^2 = 6(\dot{H} + 2H^2),
\end{equation}
where 
$H=\frac{\dot{a}}{a}$
is the Hubble parameter.

The energy-momentum tensor for the perfect fluid is considered as,
\begin{equation}
T_{ij}=(\rho +p)u_{i}u_{j}+pg_{ij}   
\end{equation}
$u_i$ is the four velocity vectors. $u_i$ satisfy the relations $u_iu^i=-1$. The energy density is $\rho$, the pressure is $p$. $T =-\rho+3p$ is the trace of the energy-momentum tensor. 
Finally, the energy-momentum tensor of FLRW is written as, 
\begin{equation}\label{cw8}
T^i_j=\texttt{diag}[-\rho,p,p,p],  
\end{equation}
To explore the cosmological implications of the proposed modified gravity framework, we adopt a specific functional form of the arbitrary function $f(R,L_m,T)$ (reference from \cite{Haghani_2021}).
\begin{equation}
    f(R,L_m,T) = R - \mu L_m T - \gamma
\label{eq:18} \end{equation}
where $\mu$ and $\gamma$ are positive arbitrary constants.
Then the partial derivatives are,
\begin{equation}
    f_R = 1,\quad f_{L_m} = -\mu T,\quad f_T = -\mu L_m
\end{equation}
we adopt $L_m =-\rho$, where $\rho$ is the energy density. By applying Eq. \eqref{eq:18} in Eq. \eqref{eq:13}, then we get the following field equation
\begin{equation}
R_{ij} - \frac{1}{2} R g_{ij} = \left(8\pi - \frac{\mu}{2} T - \mu L_m \right) T_{ij} + \left( \mu L_m^2 - \frac{\gamma}{2} \right) g_{ij}
\end{equation}
Now, using the above functional form in general modified Friedmann field equation and we assume $L_m =-\rho$, where $\rho$ is the energy density. we give the simplified equation. The first modified Friedmann field equation is, 
\begin{equation}    
 2\dot{H} + 3H^2 = \frac{\gamma}{2} - 8\pi P - \mu \rho^2 + \frac{3}{2}\mu P(P - \rho)
\label{eq:20}\end{equation}
The second modified Friedmann field equation is,
\begin{equation}
3H^2 = \frac{\gamma}{2} + 8\pi \rho + \tfrac{1}{2}\mu \rho(\rho - 3P) 
\label{eq:21}\end{equation}
In the homogeneous and isotropic FLRW background and perfect fluid matter content, the continuity equation for given modified gravity is expressed as,
\begin{equation}
\dot{\rho} + 3H(P + \rho) = \frac{\mu\rho(3\dot{P} + \dot{\rho})}{16\pi + 3\mu(\rho - P)}
\end{equation}
\section{Qualitative Study and Cosmic Dynamics of Model}\label{sec:cwfrl2}
In this section, we discover the dynamical evolution of the Universe inside the proposed model through qualitative methods. The cosmological equations are reformulated into a self-directed system of differential equations to aid in this analysis. Investigating the linear stability of the critical points corresponding to distinct cosmological situations allows us to understand the system's stability and long-term behavior. Applying the equation of state $P=\rho \omega$. The first modified Friedmann field equation [\ref{eq:20}], can be communicated in the following form,
\begin{eqnarray}
  \frac{\dot{H}}{H^2} &=& \frac{3}{2}\left[-1 + \frac{\gamma}{6H^2} - \frac{8\pi\rho\omega}{3H^2} +\frac{\mu \rho^2(3\omega^2-3\omega -2)}{6H^2}\right] \end{eqnarray}
The second modified Friedmann field equation [\ref{eq:21}] can be reformulated as follows:
\begin{eqnarray}
1 &=& \frac{\gamma}{6H^2} + \frac{8\pi\rho}{3H^2} + \frac{\mu \rho^2(1-3\omega)}{6H^2}
\label{eq:24}
\end{eqnarray}
The evolutionary stages of the Universe can be investigated by expressing the cosmological equations as an autonomous dynamical system using the dimensionless variables defined below.
\begin{equation}
x = \frac{\gamma}{6H^2}, \quad y = \frac{8\pi \rho}{3H^2}, \quad z = \frac{\mu \rho^2(1-3\omega)}{6H^2}
\label{eq:25}
\end{equation}
 Using equations [\ref{eq:24}] and [\ref{eq:25}], the constraint equation can be formulated in terms of dynamical variables.
\begin{equation}
    1=x+y+z
\label{eq:26}
\end{equation}
Equation [\ref{eq:26}] imposes a constraint that restricts the state space to three dimensions; hence, the corresponding evolution equations are coming,
\begin{eqnarray}
  x' &=& \frac{dx}{dN}=-3 x \left(-1+x-y \omega +\frac{(3 \omega ^2-3 \omega -2) z}{1-3 \omega }\right) \label{eq:27} \\ 
  y' &=&\frac{dy}{dN} = \frac{-(\omega+1)y}{2(y+2z)} \left(6y+\frac{18z(1-\omega)}{1-3 \omega}\right)-3y\left(-1+ x-y\omega +\frac{(3\omega^2-3\omega -2)z}{1-3\omega}\right) \label{eq:28} \\ 
  z' &=& \frac{dz}{dN} = \frac{-(\omega +1)z}{y+2z}\left(6y+\frac{18z(1-\omega)}{1-3 \omega}\right)-3z\left(-1+ x-y\omega +\frac{(3\omega ^2-3\omega-2)z}{1-3\omega}\right) \label{eq:29}
\end{eqnarray}
The above equations include the equation of state parameter (EOS) $P = \rho \omega $, which defines the correlation between pressure ($P$) and energy density ($\rho$) for dissimilar fluid components. Since equations (\ref{eq:27}), (\ref{eq:28}) and (\ref{eq:29}) do not contain explicit dependence on the ‘time’ parameter $N$, the system qualifies as an autonomous dynamical system. In this formulation, a prime denotes differentiation with respect to $N$, where $N = \log a$ , $a$ is the scale factor. To connect the theoretical model with observational data; it is useful to define several quantities of cosmological relevance. One such quantity is the deceleration parameter $q$, which measures the accelerating or decelerating behaviour of the universe. The deceleration parameter $q$ is expressed as,
\begin{eqnarray}
    q&=&-1-\left[\frac{3}{2} \left(x-\omega  (y+z)+\frac{8 z}{9 \omega -3}+\frac{2 z}{3}-1\right)\right] 
\end{eqnarray}
The effective equation of state parameter 
$\omega_{\text{eff}}$ is expressed in the form,
\begin{eqnarray} \omega_{\text{eff}}&=&\frac{1}{3}(2q-1)=-x+y \omega +\left(\omega +\frac{8}{3-9 \omega }-\frac{2}{3}\right) z
\end{eqnarray}

Equation of State parameter $\omega$ for different cosmic fluids are given in Table-\ref{Table I}
\begin{table}[h!]
    \centering
\begin{tabular}{|c|c|}
\hline
\textbf{Value of $\omega$} & \textbf{Cosmic Fluid} \\
\hline $1$ & Stiff matter \\
$\tfrac{1}{3}$ & Radiation \\
$0$ & Baryonic (dust-like) matter \\
$- \tfrac{1}{3}$ & Cosmic strings \\
$- \tfrac{2}{3}$ & Domain walls \\
$-1$ & Cosmological constant--like fluid \\
\hline
\end{tabular}
\caption{Equation of State parameter $\omega$ for different cosmic fluids \cite{Shukla_2025}}
\label{Table I}
\end{table}

\subsection{Phase - Space Study of the Model}
In studying the system defined by equations \eqref{eq:27} - \eqref{eq:29}, the primary task is to find the critical points. In this context, the critical points represent solutions of the differential equations where conditions $x'=0$, $y'=0$ and $z'=0$ are satisfied. The phase space behavior of the system is analyzed by studying the stability of the critical points ($x_*$, $y_*$, $z_*$), which is determined through the eigenvalues of the Jacobian matrix evaluated at these points. More information is provided in \cite{book}. The autonomous system given by equations \eqref{eq:27} - \eqref{eq:29} admits five critical points, denoted as $A, B, C, D$ and $E$. Table \ref{Table II} summarizes the corresponding cosmological parameters, the nature of these points and the eigenvalues of the Jacobian matrix evaluated at each point. In this section, we analyze the stability of the proposed model described by autonomous system of equations. The system admits five critical points and their properties are summarized in Table \ref{Table II}. A detailed discussion of these points is presented below, with emphasis on their cosmological interpretation and stability behavior.
\begin{itemize}
   \item{\textbf{Point $A$:}} 
This critical point corresponds to a formation with coordinates $\left (0, \frac{3(1-\omega)}{2}, \frac{3\omega-1}{2}\right)$. The deceleration parameter $q=-1$ and effective equation of state $\omega_{\text{eff}}=-1$ (refer Table \ref{Table II}) are fixed at this point, which represents an accelerated expansion similar to de Sitter. The eigenvalues associated with this fixed point are $\lambda_1=0$, $\lambda_2=-3$, $\lambda_3=\frac{9(\omega^2-1)}{3\omega+1}$ includes a zero eigenvalue, this fixed point is non-hyperbolic, and is normally hyperbolic \cite{AAKoley1999} showing stability at $-1<\omega <-\frac{1}{3}\lor \omega >1$, where the remaining eigenvalues shows negative behaviour. This critical point can be interpreted as describing the current phase of the Universe’s accelerated expansion, where the cosmic dynamics are dominated by an effective dark energy component driving the de Sitter–like evolution.

\item{\textbf{Point $B$:}} At this fixed point, the coordinates are $(0,1,0)$ with deceleration parameter $q=\tfrac{1}{2}(3\omega+1)$ and equation of state $\omega_{\text{eff}}=\omega$. The eigenvalues of fixed point B are $\lambda_1=3(-\omega-1),\lambda_2=3\omega,\lambda_3=3(\omega+1)$ (refer Table \ref{Table II}) are shows mixed signs, this point $B$ as classifying a saddle behaviour within the range $\omega >0$, where $\omega$ describe early time phases of Universe evolution such as matter-domintated phase at $\omega=0$, radiation dominated phase at $\omega=\frac{1}{3}$. This implies that point $B$ describes a transient phase of cosmic evolution representing the standard radiation or matter era, the system may pass near this state before evolving toward the late-time attractor.

\item{\textbf{Point C:}}
The coordinates of fixed point $C$ are $(0,0,1)$ with deceleration parameter $q=\tfrac{-9\omega^2+6\omega+7}{2-6\omega}$ and equation of state $\omega_{\text{eff}}=\tfrac{3(\omega-1)\omega-2}{3\omega-1}$ both dependent on $\omega$. Critical point $C$ can represent quintessence-like, phantom-like, or exotic intermediate regimes depending on the equation of state parameter. For many choices of $\omega$, the eigenvalues are $\lambda_1= \frac {9(\omega^2-1)}{2(3\omega-1)}, \lambda_2= \frac {9(\omega^2-1)}{3\omega-1}, \lambda_3= \frac {3(3\omega^2-3\omega-2)}{3\omega-1}$ (refer Table \ref{Table II}) include positive real parts, making $C$ a repeller or unstable saddle. This instability suggests that point $C$ is more relevant to early-time cosmology, possibly describing the departure from stiff-fluid or radiation-dominated conditions. Its role is associated with initial phases of expansion that set the stage for subsequent cosmic evolution. Moreover, this critical point is describing the decelerating phase within the range $\frac{1}{9} \left(2-\sqrt{73}\right)<\omega <\frac{1}{3}\lor \omega >\frac{1}{9} \left(\sqrt{73}+2\right)$, and is showing unstable behaviour for $\frac{1}{6} \left(3-\sqrt{33}\right)<\omega <\frac{1}{3}\lor \omega >\frac{1}{6} \left(\sqrt{33}+3\right)$, where all the eigenvalues are showing the positive signature.

\item{\textbf{Point D:}}
The proposed model includes the constant critical point $(1,0,0)$. The eigenvalues corresponding to this fixed point are $\lambda_1=-4, \lambda_2=-3, \lambda_3=-1$. It should be noted that eigen values are strictly negative. As a result this fixed point is stable. At this fixed point, universe is undergoing the phase of acceleration expansion described by the corresponding decelerator parameter $q$  having the negative value of $-1 < 0$. $a \propto e^{kt}$ is the scale factor of the universe,  where $k$ is constant. At this particular point, the effective equation of state parameter can be represented as $\omega_{\text{eff}}= -1$. Its confirming that fixed point D is a stable attractor for all values of $\omega$. Physically, this point corresponds to a de Sitter–like accelerated universe, making it the natural candidate for the late-time fate of cosmic evolution in the model. Its strong stability across different parameter choices supports the idea that the model generically evolves toward a dark-energy-dominated accelerated phase.\\

\item{\textbf{Point E:}}
At this fixed point E, the coordinates are $\left(\tfrac{3\omega+1}{3\omega - 1}, -\tfrac{3(1 - \omega)}{3\omega-1}, - 1\right)$ depend on $\omega$. Both the deceleration parameter $q = - 1$ and effective equation of state $\omega_{\text{eff}}= -1$ (refer table \ref{Table II}) have constant values, imitating a cosmological constant–like phase. Due to the presence of zero eigenvalues, this point is non-hyperbolic and its stability cannot be established solely by linear perturbation analysis. Additional methods, such as center manifold theory or numerical evolution are necessary to assess its long-term behavior. However, its effective parameters indicate that it may act as another candidate for a dark-energy-dominated solution. This critical point showa stability within the range $-1<\omega <-\frac{1}{3}\lor \omega >1$, the part of this range of EoS parameter agrees to the quintessence behaviour.
\end{itemize}

\begin{table}[h!]
    \centering
\begin{tabular}{|c|c|c|c|c|c|c|c|c|}
\hline
 \text{Point} & $x_*$ & $y_*$ & $z_*$ & $\lambda _1$ & $\lambda _2$ & $\lambda _3$ & $q$ & $\omega _{\text{eff}}$ \\
 \hline
 A & 0 & $\frac{3 (1-\omega )}{2}$ & $\frac{(3 \omega -1)}{2}$ & 0 & -3 &$ -\frac{9 \left(\omega ^2-1\right)}{3 \omega +1}$ & -1 & -1 \\
 \hline
 B & 0 & 1 & 0 & $3 (-\omega -1)$ &$ 3 \omega$  & $3 (\omega +1)$ & $\frac{1}{2} (3 \omega +1)$ & $\omega $ \\
 \hline
 C & 0 & 0 & 1 & $\frac{9 \left(\omega ^2-1\right)}{2 (3 \omega -1)}$ & $\frac{9 \left(\omega ^2-1\right)}{3 \omega -1} $& $\frac{3 \left(3 \omega ^2-3 \omega -2\right)}{3 \omega -1}$ &$ \frac{-9 \omega ^2+6 \omega +7}{2-6 \omega } $& $\frac{3 (\omega -1) \omega -2}{3 \omega -1} $\\
 \hline
 D & 1 & 0 & 0 & -4 & -3 & -2 & -1 & -1 \\
 \hline
  E & $\frac{3 \omega +1}{3 \omega -1}$ & $-\frac{3 (1-\omega )}{3 \omega -1}$ & -1 & 0 & $-\frac{9 \left(\omega ^2-1\right)}{3 \omega +1}$ & $\frac{3 \left(-27 \omega ^3+9 \omega ^2+3 \omega -1\right)}{(3 \omega -1)^2 (3 \omega +1)}$ & -1 & -1 \\
 \hline
\end{tabular}
\caption{Cosmological Phases and Model Behavior Corresponding to Critical Points}
\label{Table II}
\end{table}

\begin{figure}[H]
    \centering    \includegraphics[width=0.5\linewidth]{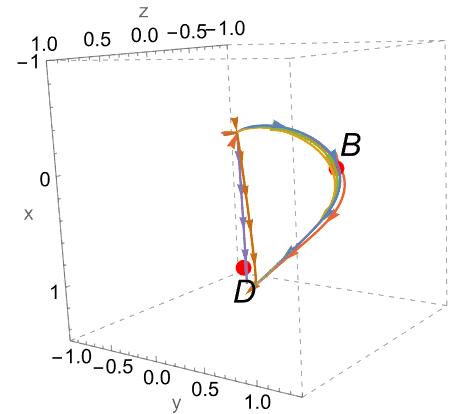}
    \caption{3D plot phase space to analyse the behaviour of phase space trajectories.}
    \label{fig:graph1}
\end{figure}
\begin{figure}[H]
    \centering
    \includegraphics[width=0.45\textwidth]{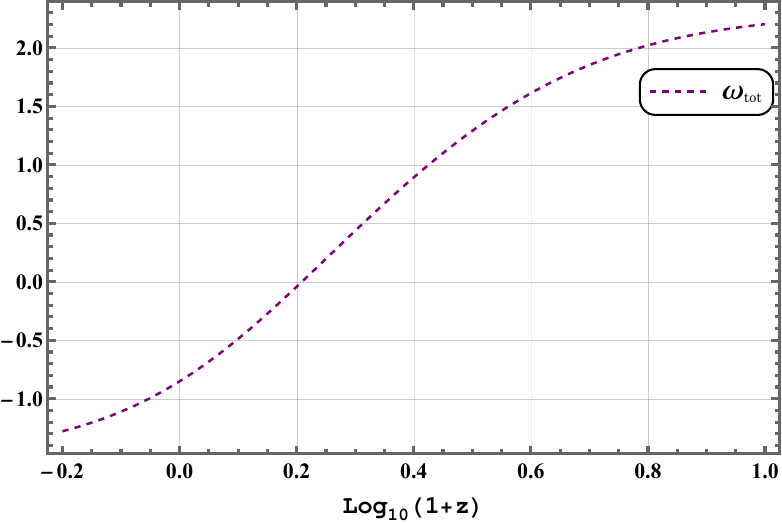} \hfill
    \includegraphics[width=0.45\textwidth]{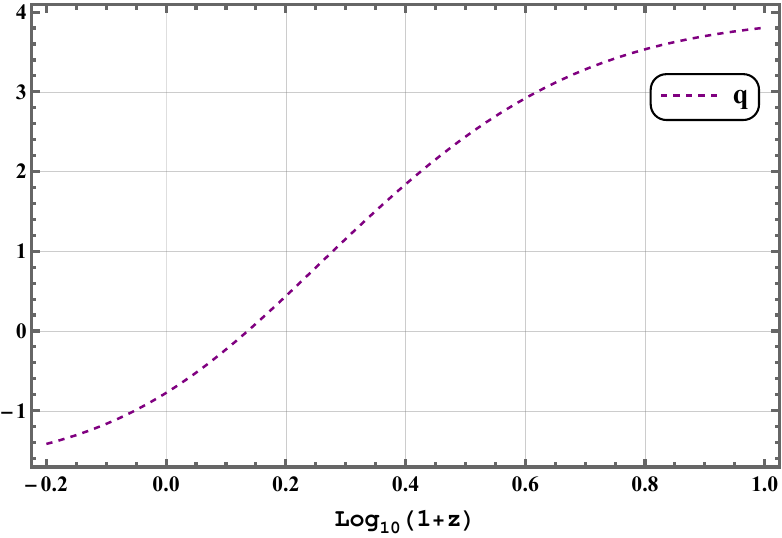}
    \caption{Behaviour of EoS and deceleration parameters, the initial conditions are $x_0=10^{-8.89},y_0=10^{-4},z_0=\sqrt{0.999}$.}
    \label{fig:Eos:deceleration}
\end{figure}
To investigate the cosmological dynamics through the dynamical parameter, effective EoS parameter have been plotted in Fig. \ref{fig:Eos:deceleration}. The same have been retested using the deceleration parameter presented in the same figure. Both the parameter describe the early time deceleration to current cosmic acceleration phenomena. The model effectively captures the present dynamics of cosmic acceleration, the EoS parameter value is \( \omega_{tot}(z=0) = -1 \pm 0.2 \), which aligns with the latest measurements from the Planck Collaboration, where \( \omega(z=0) = -1.028 \pm 0.032 \) as reported in \cite{Aghanim:2018eyx}. Additionally, the current deceleration parameter \( q(z=0) \approx -0.82 \) is in agreement with findings from contemporary observational studies \cite{PhysRevResearch.2.013028}.

\section{Conclusion}\label{conclusion}
In this work, we studied different phases of Universe evolution using the dynamical system analysis inside the framework of modified $f(R,L_{m},T)$ gravity. Specifically, we considered a matter-geometry coupling form of the model given by $f(R,L_{m},T)=R-\mu L_{m} T-\gamma$ , where $\mu$ and $\gamma$ are positive arbitrary constants. The corresponding autonomous dynamical system, derived from equations \eqref{eq:27}-\eqref{eq:29}, analyzed under the assumption of a barotropic equation of state $p =\rho \omega$. Through this dynamical systems approach, we identified the critical points of the model and evaluated their stability by computing the associated eigenvalues, as summarized in Table \ref{Table II}. The stability analysis of the autonomous system reveals that the critical points describes different phases depending on the value of the EoS $\omega$. Points $B$ and $C$ generally act as saddle and unstable (repellers) respectively and therefore can be associated with transient cosmological phases such as radiation- or matter-dominated eras. Critical points $A, D$ and $E$ remain de-Sitter solutions with $\omega=-1$, particularly critical point $D$ consistently appears as a stable attractor for all considered values of $\omega$, It is also confirmed that the value of deceleration parameter
$q =-1$ (de-Sitter Universe) at these critical points, representing the late-time accelerated phase of cosmic evolution. This behavior highlights a natural progression of the system from unstable early-time solutions toward a stable dark energy–dominated Universe can be visualised from Fig \ref{fig:graph1}. The behaviour of EoS and deceleration parameter have been investigated in Fig. \ref{fig:Eos:deceleration}, the current EoS parameter value is \( \omega_{tot}(z=0) = -1 \pm 0.2 \), which aligns with the latest measurements from the Planck Collaboration, where \( \omega(z=0) = -1.028 \pm 0.032 \) as reported in \cite{Aghanim:2018eyx}. Additionally, the current deceleration parameter \( q(z=0) \approx -0.82 \) is in agreement with findings from recent observational studies \cite{PhysRevResearch.2.013028}.

\section*{Acknowledgement}  
The author gratefully acknowledge that this research was carried out without any financial support from funding agencies in the public, commercial, or not-for-profit sectors.
\bibliographystyle{utphys}
\bibliography{biblio}

@article{Riess_1998,
   title="{Observational Evidence from Supernovae for an Accelerating Universe and a Cosmological Constant}",
   volume={116},
   ISSN={0004-6256},
   DOI={10.1086/300499},
   journal={The Astronomical Journal},
   author={Riess, Adam G. and Filippenko, Alexei V. and Challis, Peter and others},
   year={1998} 
}

@article{Perlmutter_1999,
   title="{Measurements of $\Omega$ and $\Lambda$ from 42 High‐Redshift Supernovae}",
   volume={517},
   ISSN={1538-4357},
   DOI={10.1086/307221},
   journal={The Astrophysical Journal},
   author={Perlmutter, S. and Aldering, G. and Goldhaber, G. and others},
   year={1999} 
}

@article{Carroll:1991mt,
    author = "Carroll, Sean M. and Press, William H. and Turner, Edwin L.",
    title = "{The Cosmological constant}",
    doi = "10.1146/annurev.aa.30.090192.002435",
    journal = "Ann. Rev. Astron. Astrophys.",
    volume = "30",
    year = "1992"
}

@article{Einstein:1917ce,
    author = "Einstein, Albert",
    title = "{Cosmological Considerations in the General Theory of Relativity}",
    journal = "Sitzungsber. Preuss. Akad. Wiss. Berlin (Math. Phys. )",
    year = "1917"}

@article{Peebles:2002gy,
    author = "Peebles, P. J. E. and Ratra, Bharat",
    editor = "Hsu, Jong-Ping and Fine, D.",
    title = "{The Cosmological Constant and Dark Energy}",
    doi = "10.1103/RevModPhys.75.559",
    journal = "Rev. Mod. Phys.",
    volume = "75",
    year = "2003"
}

@article{Planck:2018vyg,
    author = "Aghanim, N. and others",
    collaboration = "Planck",
    title = "{Planck 2018 results. VI. Cosmological parameters}",
   doi = "10.1051/0004-6361/201833910",
    journal = "Astron. Astrophys.",
    volume = "641",
    year = "2020",
}

@article{RevModPhys_61_1,
  title = "{The cosmological constant problem}",
  author = {Weinberg, Steven},
  journal = {Rev. Mod. Phys.},
  volume = {61},
  year = {"1989"},
  doi = {10.1103/RevModPhys.61.1}
}

@article{Bull_2016,
title = "{Beyond $\Lambda$CDM: Problems, solutions, and the road ahead}",
journal = {Physics of the Dark Universe},
volume = {12},
year = {2016},
issn = {2212-6864},
doi = {https://doi.org/10.1016/j.dark.2016.02.001},
author = "Philip Bull and Yashar Akrami and Julian Adamek and Tessa Baker and others"
}

@article{Shukla_2025,
   title="{Dynamical Systems Analysis of $f(R,Lm)$ Gravity Model}",
   ISSN={1793-6977},
   DOI={10.1142/s021988782550118x},
   journal={International Journal of Geometric Methods in Modern Physics},
   author={Shukla, Aman and Raushan, Rakesh and Chaubey, Raghavendra},
   year={2025}
}

@article{COPELAND_2006,
   title="{Dynamics of Dark Energy}",
   volume={15},
   ISSN={1793-6594},
   DOI={10.1142/s021827180600942x},
   number={11},
   journal={International Journal of Modern Physics D},
   author={Copeland, Edmund J. and Sami, M. and Tsujikawa, SHINJI},
   year={2006},
}

@article{Clifton_2012,
   title="{Modified gravity and cosmology}",
   volume={513},
   ISSN={0370-1573},
   DOI={10.1016/j.physrep.2012.01.001},
   journal={Physics Reports},
   author={Clifton, Timothy and others},
   year={2012},
}

@article{Pavon:2012qn,
    author = "Pavon, Diego and Radicella, Ninfa",
    title = "{Does the entropy of the Universe tend to a maximum?}",
    doi = "10.1007/s10714-012-1457-x",
    journal = "Gen. Rel. Grav.",
    volume = "45",
    year = "2013"
}

@article{Visser:2003vq,
  author = {Visser, Matt},
  title = {Jerk, snap and the cosmological equation of state},
  journal = {Class. Quant. Grav.},
  volume = {21},
  pages = {2603--2616},
  year = {2004},
  doi = {10.1088/0264-9381/21/11/006},
  eprint = {gr-qc/0309109},
  archivePrefix = {arXiv}
}

@Article{PhysRevResearch.2.013028,
  author        = {Camarena, David and Marra, Valerio},
  journal       = {Phys. Rev. Res.},
  title         = {Local determination of the Hubble constant and the deceleration parameter},
  year          = {2020},
  pages         = {013028},
  volume        = {2},
  archiveprefix = {arXiv},
  doi           = {10.1103/PhysRevResearch.2.013028},
  eprint        = {1906.11814},
  primaryclass  = {astro-ph.CO},
}

@Article{Aghanim:2018eyx,
  author        = {Aghanim, N. and others},
  journal       = {Astron. Astrophys.},
  title         = {{Planck 2018 results. VI. Cosmological parameters}},
  year          = {2020},
  note          = {[Erratum: Astron.Astrophys. 652, C4 (2021)]},
  pages         = {A6},
  volume        = {641},
  archiveprefix = {arXiv},
  collaboration = {Planck},
  doi           = {10.1051/0004-6361/201833910},
  eprint        = {1807.06209},
  primaryclass  = {astro-ph.CO},
}

@article{AAKoley1999,
  url = {https://arxiv.org/abs/gr-qc/9910074},
  author = {Coley, A. A.},
  title = {Dynamical Systems in Cosmology},
  year = {1999},
  }

@article{Zhang:2005yz,
  author = {Zhang, Xin},
  title = {Statefinder diagnostic for holographic dark energy model},
  journal = {Int. J. Mod. Phys. D},
  volume = {14},
  pages = {1597--1606},
  year = {2005},
  doi = {10.1142/S0218271805007243},
  eprint = {astro-ph/0504586},
  archivePrefix = {arXiv}
}

@article{Nojiri:2006ri,
  author = {Nojiri, Shin'ichi and Odintsov, Sergei D.},
  title = {Introduction to modified gravity and gravitational alternative for dark energy},
  journal = {Int. J. Geom. Meth. Mod. Phys.},
  volume = {4},
  pages = {115--146},
  year = {2007},
  doi = {10.1142/S0219887807001928},
  eprint = {hep-th/0601213},
  archivePrefix = {arXiv}
}

@article{Lusso:2016vmu,
  author = {Lusso, Elisabeta and Risaliti, Guido},
  title = {A Hubble Diagram for Quasars},
  journal = {Astrophys. J.},
  volume = {819},
  number = {2},
  pages = {154},
  year = {2016},
  doi = {10.3847/0004-637X/819/2/154},
  eprint = {1602.01077},
  archivePrefix = {arXiv}
}

@article{Harko_2011,
   title="{$f(R,T)$ gravity}",
   volume={84},
   ISSN={1550-2368},
   DOI={10.1103/physrevd.84.024020},
   journal={Physical Review D},
   author={Harko, Tiberiu and Lobo, Francisco S. N. and Nojiri, Shin’ichi and Odintsov, Sergei D.},
   year={2011},
}

@article{Tegmark_2004,
   title="{Cosmological parameters from SDSS and WMAP}",
   volume={69},
   ISSN={1550-2368},
   DOI={10.1103/physrevd.69.103501},
   journal={Physical Review D},
   author={Tegmark, Max and Strauss, Michael A. and others},
   year={2004}
}

@article{Harko_2010,
   title="{$f(R,L_m)$ gravity}",
   volume={70},
   ISSN={1434-6052},
   DOI={10.1140/epjc/s10052-010-1467-3},
   journal={The European Physical Journal C},
   author={Harko, Tiberiu and Lobo, Francisco S. N.},
   year={2010},
}

@article{Sotiriou_2010,
   title="{$f(R)$ theories of gravity}",
   volume={82},
   ISSN={1539-0756},
   DOI={10.1103/revmodphys.82.451},
   journal={Reviews of Modern Physics},
   author={Sotiriou, Thomas P. and Faraoni, Valerio},
   year={2010},
 }

@article{De_Felice_2010,
   title="{$f(R)$ Theories}",
   volume={13},
   ISSN={1433-8351},
   DOI={10.12942/lrr-2010-3},
   journal={Living Reviews in Relativity},
   author={De Felice, Antonio and Tsujikawa, Shinji},
   year={2010},
}

@article{Harko_2012,
   title="{Metric-Palatini gravity unifying local constraints and late-time cosmic acceleration}",
   volume={85},
   ISSN={1550-2368},
   DOI={10.1103/physrevd.85.084016},
   journal={Physical Review D},
   author={Harko, Tiberiu and others},
   year={2012},
}

@article{Capozziello_2015,
   title="{Hybrid Metric-Palatini Gravity}",
   volume={1},
   ISSN={2218-1997},
   DOI={10.3390/universe1020199},
   journal={Universe},
   author={Capozziello, Salvatore and Harko, Tiberiu and Koivisto, Tomi and Lobo, Francisco and Olmo, Gonzalo},
   year={2015},
}

@article{Capozziello_2011,
   title="{Extended Theories of Gravity}",
   volume={509},
   ISSN={0370-1573},
   DOI={10.1016/j.physrep.2011.09.003},
   journal={Physics Reports},
   author={Capozziello, Salvatore and De Laurentis, Mariafelicia},
   year={2011},
 }

@article{Carloni_2015,
   title="{Dynamical system analysis of hybrid metric-Palatini cosmologies}",
   volume={92},
   ISSN={1550-2368},
   DOI={10.1103/physrevd.92.064035},
   journal={Physical Review D},
   author={Carloni, Sante and Koivisto, Tomi and Lobo, Francisco S. N.},
   year={2015},
}

@article{Haghani_2021,
   title="{Generalizing the coupling between geometry and matter: $f\left(R,L_m,T\right)$ gravity}",
   volume={81},
   ISSN={1434-6052},
   DOI={10.1140/epjc/s10052-021-09359-3},
   journal={The European Physical Journal C},
   author={Haghani, Zahra and Harko, Tiberiu},
   year={2021},
}

@book{LandauLifshitz1980,
  author       = {Landau, Lev D. and Lifshitz, Evgenii M.},
  title        = "{The Classical Theory of Fields}",
  volume       = {2},
  edition      = {4},
  year         = {1980},
  publisher    = {Butterworth-Heinemann},
  address      = {Oxford},
  isbn         = {9780750627689},
  isbnebook    = {9780080503493},
}

@book{Ryden:1970vsj,
    author = "Ryden, B.",
    title = "{Introduction to cosmology}",
    doi = "10.1017/9781316651087",
    isbn = "978-1-107-15483-4, 978-1-316-88984-8, 978-1-316-65108-7",
    publisher = "Cambridge University Press",
    year = "1970"
}

@book{book,
    author = "Wiggins, Stephen",
    year = "2003",
    title = "Introduction To Applied Nonlinear Dynamical Systems And Chaos",
    volume = {4},
    isbn = {0-387-00177-8},
    journal = {Computers in Physics},
    doi = {10.1007/b97481}
}

@article{Moraes_2024,
   title="{Wormholes in the f(R,L,T) theory of gravity}",
   volume={855},
   ISSN={0370-2693},
   DOI={10.1016/j.physletb.2024.138818},
   journal={Physics Letters B},
   author={Moraes, P.H.R.S. and Agrawal, A.S. and Mishra, B.},
   year={2024},
 }

@article{Harko_2014,
   title="{Generalized Curvature-Matter Couplings in Modified Gravity}",
   volume={2},
   ISSN={2075-4434},
   DOI={10.3390/galaxies2030410},
   journal={Galaxies},
   author={Harko, Tiberiu and Lobo, Francisco},
   year={2014},
 }

@article{Nojiri_2011,
   title="{Unified cosmic history in modified gravity: From F(R) theory to Lorentz non-invariant models}",
   volume={505},
   ISSN={0370-1573},
   DOI={10.1016/j.physrep.2011.04.001},
   journal={Physics Reports},
   author={Nojiri, Shin’ichi and Odintsov, Sergei D.},
   year={2011},
 }

@article{Bamba_2012,
   title="{Dark energy cosmology: the equivalent description via different theoretical models and cosmography tests}",
   volume={342},
   ISSN={1572-946X},
   DOI={10.1007/s10509-012-1181-8},
   journal={Astrophysics and Space Science},
   author={Bamba, Kazuharu and Capozziello, Salvatore and Nojiri, Shin’ichi and Odintsov, Sergei D.},
   year={2012},
 }

@article{Fortunato:2024ulg,
    author = "Fortunato, J. A. S. and others",
    title = "{Hydrostatic equilibrium configurations of neutron stars in the f(R,L,T) gravity theory}",
    doi = "10.1016/j.dark.2025.101893",
    journal = "Phys. Dark Univ.",
    year = "2025"
}

@article{Mota_2024,
   title="{Neutron stars in $f(R,L_m,T)$ gravity}",
   volume={84},
   ISSN={1434-6052},
   DOI={10.1140/epjc/s10052-024-13042-8},
   journal={The European Physical Journal C},
   author={Mota, Clésio E. and Pretel, Juan M. Z. and Flores, César O. V.},
   year={2024}
 }

@article{Loewer_2024,
   title="{A study of stable wormhole solution with non-commutative geometry in the framework of linear $f(R,{\mathcal {L}}_m, T)$ gravity}",
   volume={84},
   ISSN={1434-6052},  
   DOI={10.1140/epjc/s10052-024-13604-w},
   journal={The European Physical Journal C},
   author={Loewer, Niklas and Tayde, Moreshwar and Sahoo, P. K.},
   year={2024},
 }

\end{document}